\documentclass[fleqn,usenatbib]{mnras}

\usepackage[T1]{fontenc}

\DeclareRobustCommand{\VAN}[3]{#2}
\let\VANthebibliography\thebibliography
\def\thebibliography{\DeclareRobustCommand{\VAN}[3]{##3}\VANthebibliography}

\usepackage{graphicx}	% Including figure files
\usepackage{amsmath}	% Advanced maths commands
\usepackage{amssymb}	% Extra maths symbols

\newcommand{\deriv}[2]{\frac{{\mathrm d} #1}{{\mathrm d} #2}}

\newcommand{\pder}[2]{ \frac{\partial #1}{\partial #2} }
\newcommand{\pderN}[3]{ \frac{\partial^{#3} #1}{\partial #2^{#3}} }

\title{On collective nature of nonlinear torsional Alfv\'en waves}

\author[S.A. Belov et al.]{S.A. Belov$^{1,2}$\thanks{E-mail: mr\_beloff@mail.ru},
D.I. Riashchikov$^{1,2}$,
D.Y. Kolotkov$^{3}$,
S. Vasheghani Farahani$^{4}$,
N.E. Molevich$^{1,2}$,
\newauthor
V.V. Bezrukovs$^{3}$
\\
$^{1}$Department of Theoretical Physics, Lebedev Physical Institute, Novo-Sadovaya st. 221, Samara, 443011, Russia\\
$^{2}$Department of Physics, Samara National Research University, Moscovskoe sh. 34, Samara, 443086, Russia\\
$^{3}$Engineering Research Institute \lq\lq Ventspils International Radio Astronomy Centre (VIRAC)\rq\rq\ of Ventspils University of Applied Sciences,\\ Inzenieru iela 101, Ventspils, LV-3601, Latvia\\
$^{4}$Department of Physics, Tafresh University, Tafresh 39518 79611, Iran
}

\date{Accepted XXX. Received YYY; in original form ZZZ}

\pubyear{2023}

\begin{document}
\label{firstpage}
\pagerange{\pageref{firstpage}--\pageref{lastpage}}
\maketitle

\begin{abstract}
 Torsional Alfv\'en waves in coronal plasma loops are usually considered to be non-collective, i.e. consist of cylindrical surfaces evolving independently, which significantly complicates their detection in observations. This non-collective nature, however, can get modified in the nonlinear regime. 
{To address this question, the propagation of nonlinear torsional Alfv\'en waves in straight magnetic flux tubes has been investigated numerically using the astrophysical MHD code Athena++ and analytically,  to support numerical results, using the perturbation theory up to the second order. Numerical results have revealed that there is radially uniform induced density perturbation whose uniformity does not depend on the radial structure of the mother Alfv\'en wave.  Our analysis showed that the ponderomotive force leads to the induction of the radial and axial velocity perturbations, while the mechanism for the density perturbation is provided by a non-equal elasticity of a magnetic flux tube in the radial and axial directions.
{The latter can be qualitatively understood by the interplay between the Alfv\'en wave perturbations, external medium, and the flux tube boundary conditions.}
The amplitude of these nonlinearly induced density perturbations is found to be determined by the amplitude of the Alfv\'en driver squared and the plasma parameter $\beta$.
The existence of the collective and radially uniform density perturbation accompanying nonlinear torsional Alfv\'en waves could be considered as an additional observational signature of Alfv\'en waves in the upper layers of the solar atmosphere. } 

\end{abstract}


%%%%%%%%%%%%%%%%%%%%%%%%%%%%%%%%%%%%%%%%%%%%%%%%%%

%%%%%%%%%%%%%%%%% BODY OF PAPER %%%%%%%%%%%%%%%%%%

\section{Introduction}
Due to the progress in observational techniques and facilities, various modes of magnetohydrodynamic (MHD) waves are being confidently detected in the solar atmosphere. In particular, the existence of upwardly propagating torsional Alfv\'en waves have been reported in photospheric bright points \citep{2009Sci...323.1582J} and thin chromospheric spicular-type structures  \citep[e.g.][]{DePontieu2014, Srivastava2017}. Likewise, there are sporadic reports of torsional Alfv\'en waves in coronal structures \citep{Kohutova2020} and of their excitation during solar flares \citep{2020ApJ...891...99A}.

 Alfv\'en waves are well-known candidates for transferring and transporting energy into the solar corona. For example, in the observation made by \citet{Srivastava2017}, the estimated energy flux carried by Alfv\'en waves at the chromospheric layer was reported to be potentially sufficient to compensate radiative losses in the corona \citep[for review of coronal heating mechanisms by Alfv\'en waves, see e.g.][]{VanDoorsselaere2020}. This is besides the great interest in Alfv\'en waves in the context of providing energy sources for solar wind acceleration \citep{Banerjee2021}.

 The evolution of torsional Alfv\'en waves in initially uniform plasmas has been shown to lead to the formation of elongated over-dense threads in high-resolution 3D ideal MHD simulations, which in turn could act as effective waveguides for other MHD wave modes \citep{Diaz-Suarez_2021}.

Despite the persistent interest in Alfv\'en waves in the solar atmosphere, there are only a few reports of their detection in the corona \citep[e.g.][]{Kohutova2020}. This is in contrast to magnetoacoustic (MA) waves which are confidently detected in the corona in direct imaging, spectroscopic, and indirect observations as quasi-periodic pulsations in flares \citep[see e.g.][for a recent comprehensive review]{Nakariakov2020}. The reason for this fact is that the detection of Aflv\'en waves is more challenging in comparison with MA waves \citep[it is also instructive to read][in this respect]{1999JPlPh..62..219V}. First of all, Alfv\'en waves are not compressive in the linear regime and, therefore, do not modulate plasma emission through density perturbations as essentially compressive MA waves do. As a result, if the plasma structure is poorly resolved, the only way to detect linear (with relatively low amplitude) Alfv\'en wave is to measure non-thermal Doppler broadenings of spectral lines. Analyzing data  of periodic non-thermal variations of line widths in coronal holes and facular regions in a number of lines, \citet{Chelpanov2022} considered that these Doppler broadenings should be necessary but not sufficient for unambiguous identification of torsional Alfv\'en waves in the lower solar atmosphere.  Indeed, for linear Alfv\'en waves, these broadenings should not be accompanied by the correlating intensity variations, which are more likely to arise due to MA perturbations. A more robust torsional Alfv\'en wave observation was made by \cite{2009Sci...323.1582J} when the detected periodic modulation of spectral line nonthermal broadening was not accompanied by cospatial intensity oscillations and transversal displacements.

The requirement of absence of the accompanying intensity modulation is relaxed in the nonlinear case. Indeed, nonlinear Alfv\'en waves can induce compressive perturbations propagating at speeds equal to that of the mother Alfv\'en wave \citep{Hollweg1971,2011A&A...526A..80V}. In this case, plasma intensity modulation due to the induced compressive perturbations should accompany Doppler line broadening. However, it is not clear, whether this intensity modulation would differ from the case of MA waves or not. The fact of the matter is that torsional Alfv\'en waves in the linear regime do not initiate a collective process. This means that in the linear regime all cylindrical surfaces hosting oscillations evolve independently \citep{2010A&A...517A..29V,2017A&A...599A..19V}. For radially non-uniform profiles of the Alfv\'en speed in a wave-hosting loop, such a non-collectivity results in phase-mixing when the wavefront distorts and smaller spatial scales arise \citep{Heyvaerts1983, Petrukhin2017, Ruderman2017, Ruderman2018, Guo2019}. Moreover, during phase-mixing, an effective nonlinear excitation of fast waves propagating across the magnetic field, away from the layer of phase mixing was reported \citep{Nakariakov1997}. {It was possible to decouple the Alfv\'en and fast modes in the latter study due to a 2.5D geometry, which may be less obvious in full 3D inhomogeneous MHD.} A similar result was obtained for the nonlinear interaction between Alfv\'enic perturbations and the solar wind current sheet, resulting in the generation of compressive perturbations \citep{Malara1996}.

In uniform loops, i.e. in the absence of phase-mixing, the non-collective nature of torsional Alfv\'en waves would also lead to the independence of perturbed cylindrical surfaces. This means that torsionally perturbed cylindrical layers do not interact with other layers of the plasma tube. Taking this into account, \citet{Kolotkov2018} excluded torsional Alfv\'en waves as a reason for the observed quasi-periodic modulation of the radio emission from the solar corona, though the observed speed was about the local Alfv\'en speed. Nonetheless, although torsional Alfv\'en waves are non-collective, the question about collectivity/non-collectivity of the plasma density perturbations that they induce in the nonlinear regime remains open.
In numerical simulations of torsional Alfv\'en waves, \citet{Shestov2017} obtained that the nonlinearly induced perturbation of density is {apparently} radially uniform, while induced perturbations of other loop parameters have radial structures prescribed by the radial structure of the mother Alfv\'en wave {(see their Fig.~2)}. However, the dependence of the obtained apparent collectivity (via radial uniformity of induced density perturbations) and non-collectivity (via non-uniformity of other loop parameters) of nonlinear torsional Alfv\'en waves on the choice of the Alfv\'en driver's radial structure has not been investigated.

In this paper, we demonstrate that the loop density perturbations induced by nonlinear torsional Alfv\'en waves remain uniform in the radial direction (i.e. collective) for both the Alfv\'en driver perturbing the entire volume of the magnetic flux tube and the Alfv\'en driver localised in a narrow annulus inside the tube.

In Section \ref{s:Model}, we present the model and numerical setup considered. In Section \ref{s:Num_res}, we describe the obtained numerical results for the propagation of nonlinear Alfv\'en waves for both types of the driver. In Section \ref{s:Analytics}, we provide analytical interpretation for the results obtained numerically in Section \ref{s:Num_res}. Finally, in Section \ref{s:Conclusions}, we summarize our conclusions and highlight future possible directions stimulated by this study.

\section{Model and numerical setup}{\label{s:Model}}

To investigate the collective/non-collective nature of compressive perturbations induced by torsional Alfv\'en waves propagating inside magnetic flux tubes, we solve the set of ideal dimensionless MHD equations numerically using the astrophysical MHD code Athena++ \citep{White2016, Felker2018, Stone2020} {implementing higher-order Godunov methods}. We should stress that the numerical experiments carried out in the present study aim to shed light on the basic features of nonlinear effects connected with the torsional Alfv\'en waves rather than find how plasma acts under specific set of physical parameters.

The numerical setup of the present study is a follow up of the work by  \citet{Shestov2017}. Thus, we consider a magnetic cylinder, where only axisymmetric motions can exist. The steady profiles of the axial magnetic field $B_0(R)$ and density $\rho_0(R)$ are given by
\begin{equation}
	B_0\left(R\right)=B_{0_i}\,\sqrt[]{1+2\frac{\left(\rho_{0_i}-\rho_{0_e}\right) T_0}{B_{0_i}^2}\left(1-S_0\left(R\right)\right)},
	\label{b_profile}
\end{equation}
\begin{equation}
	\rho_0\left(R\right)=\rho_{0_e}+\left(\rho_{0_i}-\rho_{0_e}\right)S_0\left(R\right),
	\label{rho_profile}
\end{equation}
where $S_0\left(R\right)=\left(\cosh R^\alpha\right)^{-2}$,
%\begin{equation*}
%	S_0\left(R\right)=\left(\cosh R^\alpha\right)^{-2},
%\end{equation*}
$B_{0_i}=3.16228$ is the steady dimensionless axial magnetic field inside the magnetic flux tube (we note that the dimensionless magnetic field in Athena++ is divided by a factor of $\sqrt{4 \pi}$, so that the Alfv\'en speed $C_A = B/\sqrt{\rho}$), $\rho_{0_i}=1$ and $\rho_{0_e}=0.2$ are steady dimensionless densities inside and outside the magnetic flux tube, respectively, $R = r/R_0$ is the radial coordinate normalized to the radius of the magnetic flux tube, $R_0$. The temperature $T_0 = 1$ is the same inside and outside the flux tube. Note that in this formulation of the problem the dimensionless speed of sound $C_{S_{i}} \approx 1.29$ and dimensionless Alfv\'en speed inside the tube $C_{A_{i}} \approx 3.16$. In physical units and in the context of the present study, the magnetic flux tube radius corresponds to $R_0 = 1$\,Mm,  while we take $B_{0_i} = 5.9$\,G (plasma-$\beta$ is about 0.2), $T_0 = 1$\,MK, with internal and external densities $\rho_{0_i}$ and $\rho_{0_e}$ calculated based on the electron number density inside $n_{0_i} = 10^9$\,cm$^{-3}$ and outside $n_{0_e} = 2 \times 10^8$\,cm$^{-3}$ the magnetic tube. The unit time in numerical simulations is defined as $t_0 = R_0/\sqrt{k_B T_0/m} \approx 8.5$\,s. The function $S_0\left(R\right)$ is the smoothing factor which is implemented to smooth out the sharp tube boundary, with the parameter $\alpha=20$.

\begin{figure*}
\centering
	\includegraphics[scale= 0.6]{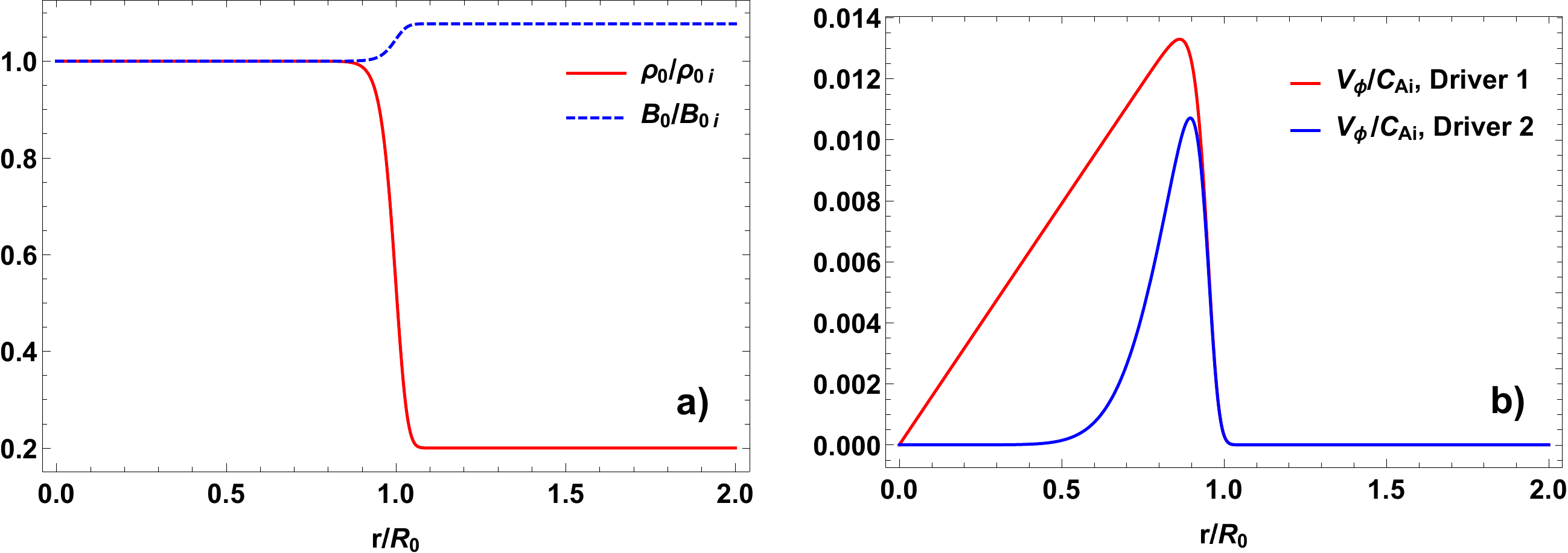}
	\caption{\textbf{a)} Initial profiles of steady axial magnetic field (blue curve) and density (red curve), given by Eqs.~(\ref{b_profile}) and (\ref{rho_profile}), respectively. \textbf{b)} Radial profiles of the Alfv\'en wave drivers, given by Eq.~(\ref{v_driver}).}
	\label{fig:setup}
\end{figure*}

The steady profiles of axial magnetic field and density which are respectively prescribed by Eqs.~(\ref{b_profile}) and (\ref{rho_profile}) are shown in Fig. \ref{fig:setup}.a.
To initiate oscillations, the drivers for torsional Alfv\'en waves are applied at the bottom of the flux tube as
\begin{equation}
	v_{\phi\,1,2}=A_0 R\sin\left(\omega t\right)S_{1,2}\left(R\right),
	\label{v_driver}
\end{equation}
\begin{equation}
	B_{\phi\,1,2}=-\sqrt{\rho_{0_i}}A_0 R\sin\left(\omega t\right)S_{1,2}\left(R\right),
	\label{b_driver}
\end{equation}
with
\begin{equation*}
	S_1\left(R\right)=\cosh\left[\left(R+0.05\right)^\alpha\right]^{-2},
	\label{s1}
\end{equation*}
\begin{equation*}
	S_2\left(R\right)=\exp\left[-\left(\frac{R-1}{0.25}\right)^2\right]S_1\left(R\right).
	\label{s2}
\end{equation*}
Here, the factor $S_1\left(R\right)$ corresponds to the Alfv\'en driver linearly increasing with radius, referred to as Driver 1 in the present study, (see the red curve in Fig.~\ref{fig:setup}.b). The factor $S_2\left(R\right)$ corresponds to the Alfv\'en driver which represents a perturbation of a narrow annulus near the magnetic flux tube boundary, called Driver 2 (see the blue curve in Fig.~\ref{fig:setup}.b). We set $A_0=0.05$ which gives the driver amplitude $A_0 R_0 \approx 0.01C_{A_{i}}$, to avoid wave steepening and other higher-order nonlinear effects \citep[see e.g.][regarding Alfv\'en wave shocks and nonlinear cascading in structured plasmas]{2012A&A...544A.127V, 2020NatSR..1015603S, 2019ApJ...882...50M, Farahani2021}. We note that in the present study the value for the dimensionless frequency $\omega$ is selected equal to $2$ (i.e. the oscillation period about 30~s).
Our choice of Drivers 1 and 2 is motivated by the aim of the present study, which is to investigate how compressive perturbations are induced in the case when the Alfv\'en wave exists in the entire volume of a magnetic flux tube (Driver 1) and when it is highly localised in the radial direction (Driver 2).

{
For both drivers, numerical simulations were carried out using the following Athena++ settings. The set of MHD equations was solved using the HLLD Riemann solver. For integration over time, a van Leer predictor-corrector method was utilized. The spatial reconstruction was performed using the piecewise linear method. This ensures second-order accuracy in time and up to second-order accuracy in space.
}

{
The numerical results outlined in Sec.~\ref{s:Num_res} were obtained in cylindrically symmetric geometry on a $800 \times 3200$ grid along the radial and axial directions, respectively, with a constant spatial resolution of $\Delta R = 0.0025 R_0$ and $\Delta z = 0.025 R_0$. That is, the mesh refinement was not applied. During the study, calculations on coarser grids were also performed, and the convergence of the results was checked. The perturbations are set with the aforementioned drivers at the bottom boundary. Outflow (zero-gradient) boundary conditions were set at the outer radial boundary and the upper boundary of the computational domain.
}

{
The simulations were conducted in two stages. First, we set up a flux tube based on the aforementioned analytical formulas and waited for the numerical setup to relax to equilibrium. In the second stage, a driver was activated to initiate perturbations.
}

%The result of these two cases is presented in the following Section \ref{s:Num_res}. 

\section{Numerical results}\label{s:Num_res}

\subsection{\lq\lq Linear\rq\rq\ driver (Driver 1)}
\begin{figure*}
	\includegraphics[width=\textwidth]{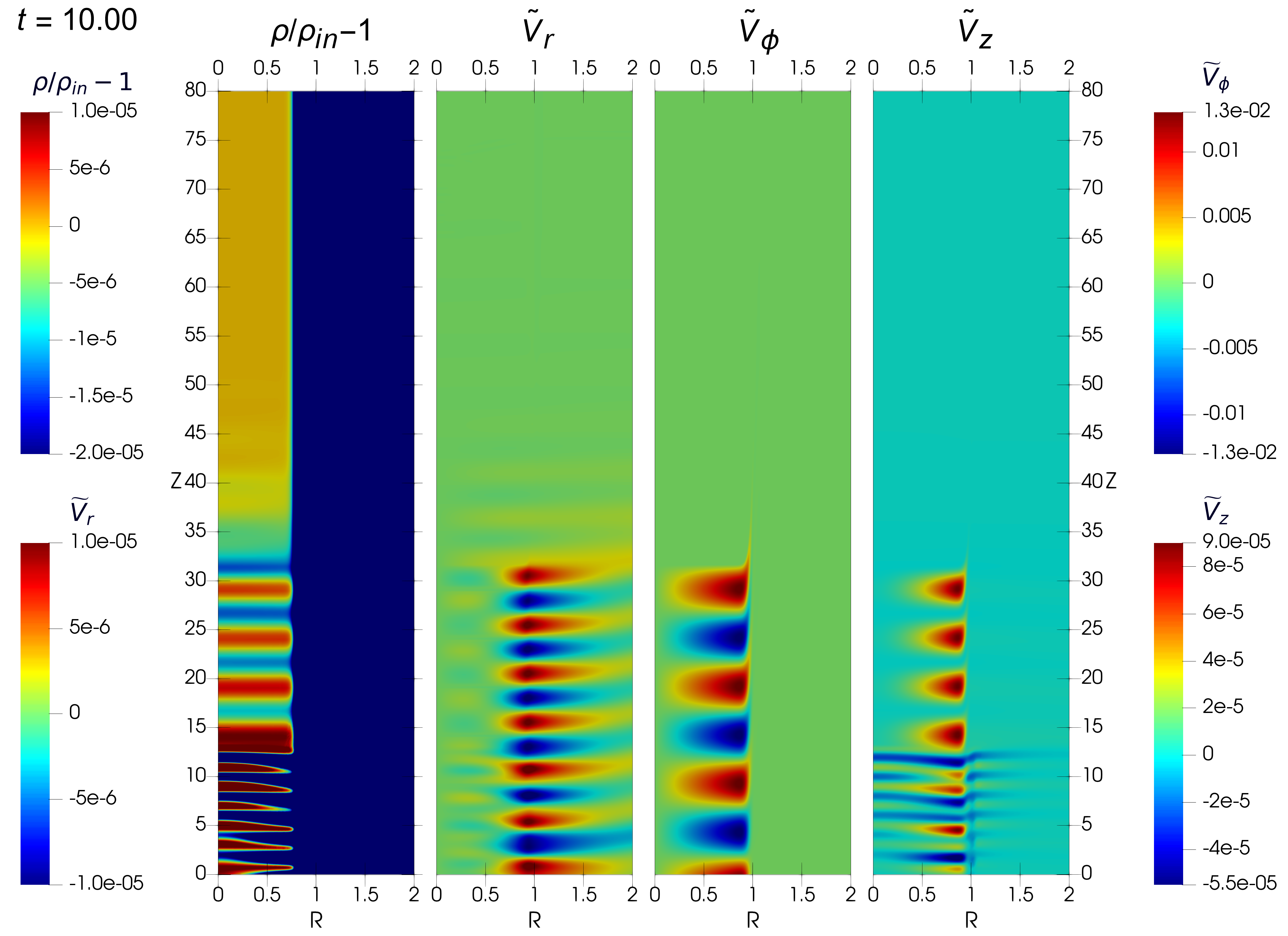}
	\caption{Dimensionless perturbations of density, axial and radial velocities induced by a torsional Alfv\'en wave (perturbation of the azimuthal velocity) existing in the whole volume of the waveguide (i.e. with Driver 1, see Fig.~\ref{fig:setup}.b) at dimensionless time $t=10$. Due to axial symmetry, only one half of the magnetic flux tube is shown. The full animation can be seen in supplementary material.}
	\label{fig:linear_full}
\end{figure*}
The results of our numerical simulations of propagating torsional Alfv\'en waves in a cylindrical magnetic flux tube are presented in Fig. \ref{fig:linear_full}, where due to axial symmetry, only one half of the magnetic flux tube is shown. In this simulation, the Alfv\'en driver is determined by Eqs. (\ref{v_driver}) and (\ref{b_driver}) with subscript 1, i.e. Driver 1. This driver corresponds to the case when the mother Alfv\'en wave exists in the entire volume of the magnetic flux tube. The radial profile of this driver is plotted in red in Fig. \ref{fig:setup}.b. The perturbation of the azimuthal velocity $\tilde{v}_\phi$ corresponds to the propagating mother torsional Alfv\'en wave. In the vicinity of the tube boundary ($R=1$), we can clearly see the development of phase-mixing effects because of the presence of the local Alfv\'en speed gradient. It can be seen from Fig. \ref{fig:linear_full} that the mother Alfv\'en wave induces perturbations of density, axial and radial velocities. These perturbations propagate at the Alfv\'en speed inside the flux tube. On plots for density and axial velocity, one can notice propagating slow waves in the lower part of the tube. Moreover, fast waves can also be seen on the radial velocity plot above $z\approx 31.6$. It should be noted that the results of our simulation agree with the results presented by \citet{Shestov2017}.

\begin{figure*}
	\includegraphics[width= \textwidth]{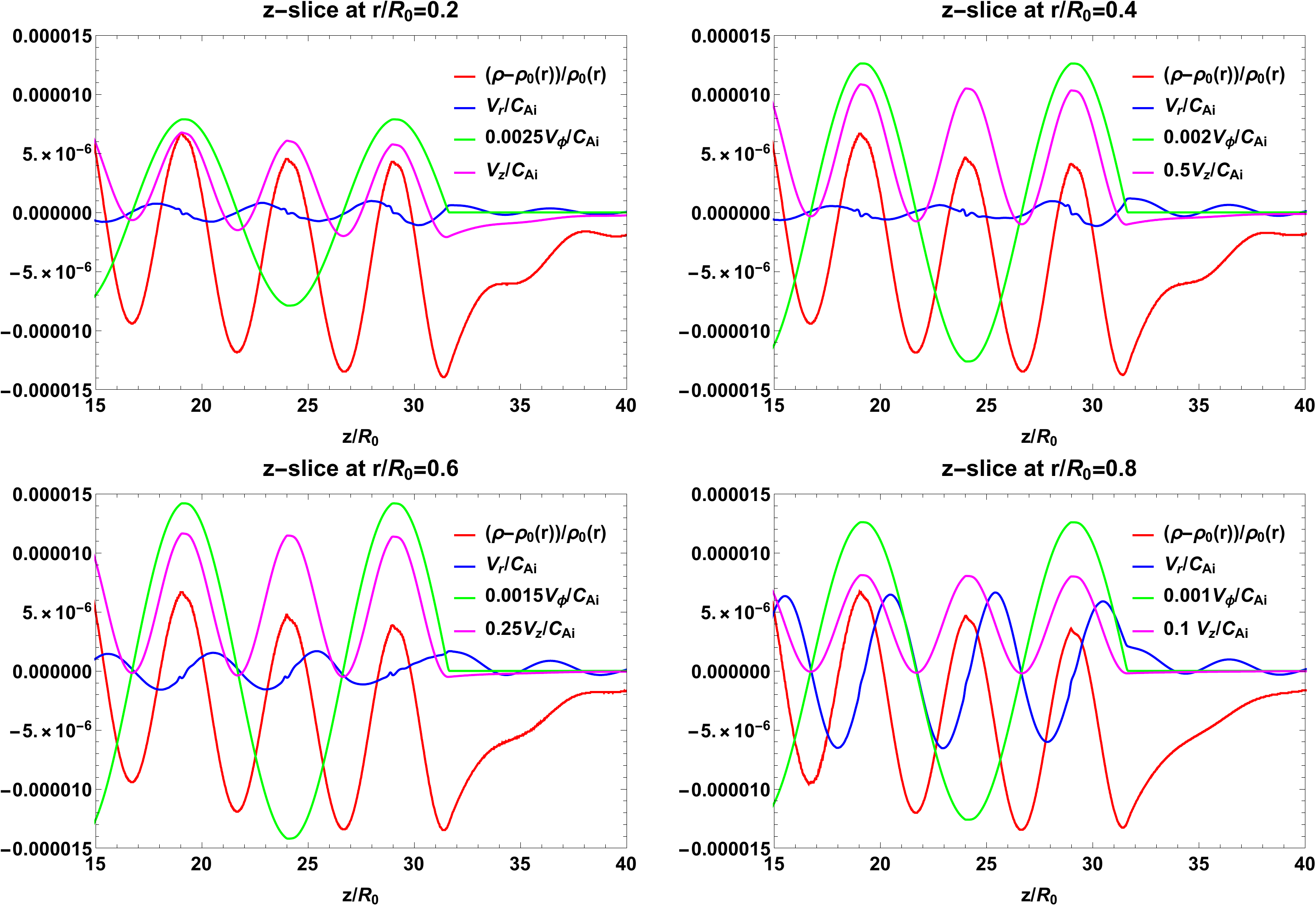}
	\caption{Plasma parameters sliced along the $z$-axis for different values of the radial coordinate, perturbed by the mother Alfv\'en wave driven by Driver 1 (see Fig.~\ref{fig:linear_full}) at dimensionless time $t=10$. Some parameters have been re-scaled for visualizing purposes.}
	\label{fig:ld_z_slices}
\end{figure*}

\begin{figure*}
	\includegraphics[width= \textwidth]{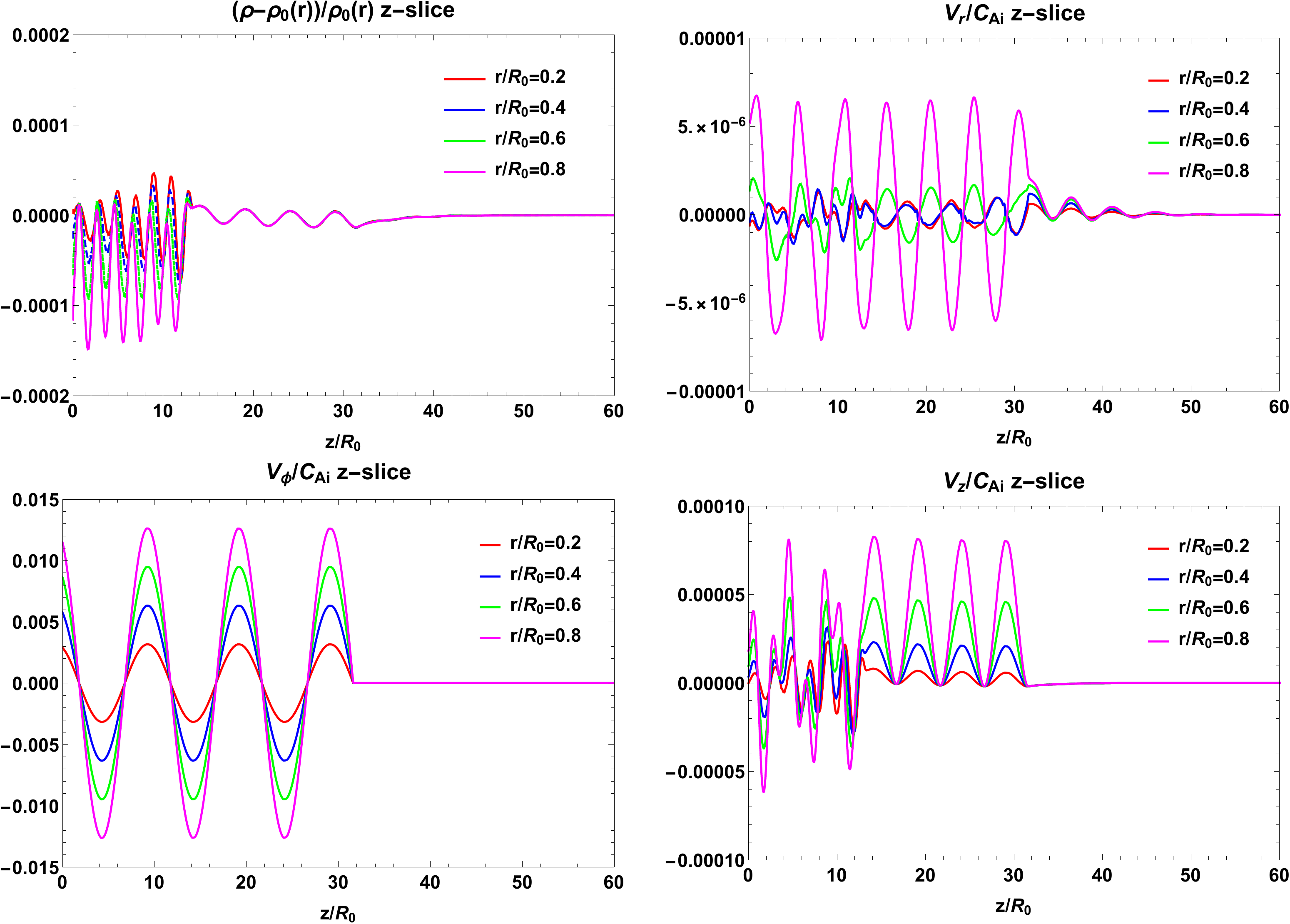}
	\caption{Comparison of the plasma parameters sliced along the $z$-axis for different values of the radial coordinate (for Driver 1) at dimensionless time $t=10$. Each plot compares axial slices of one parameter, taken at different values of the radial coordinate.}
	\label{fig:ld_separate_z_slices}
\end{figure*}

\begin{figure}
	\includegraphics[width= \columnwidth]{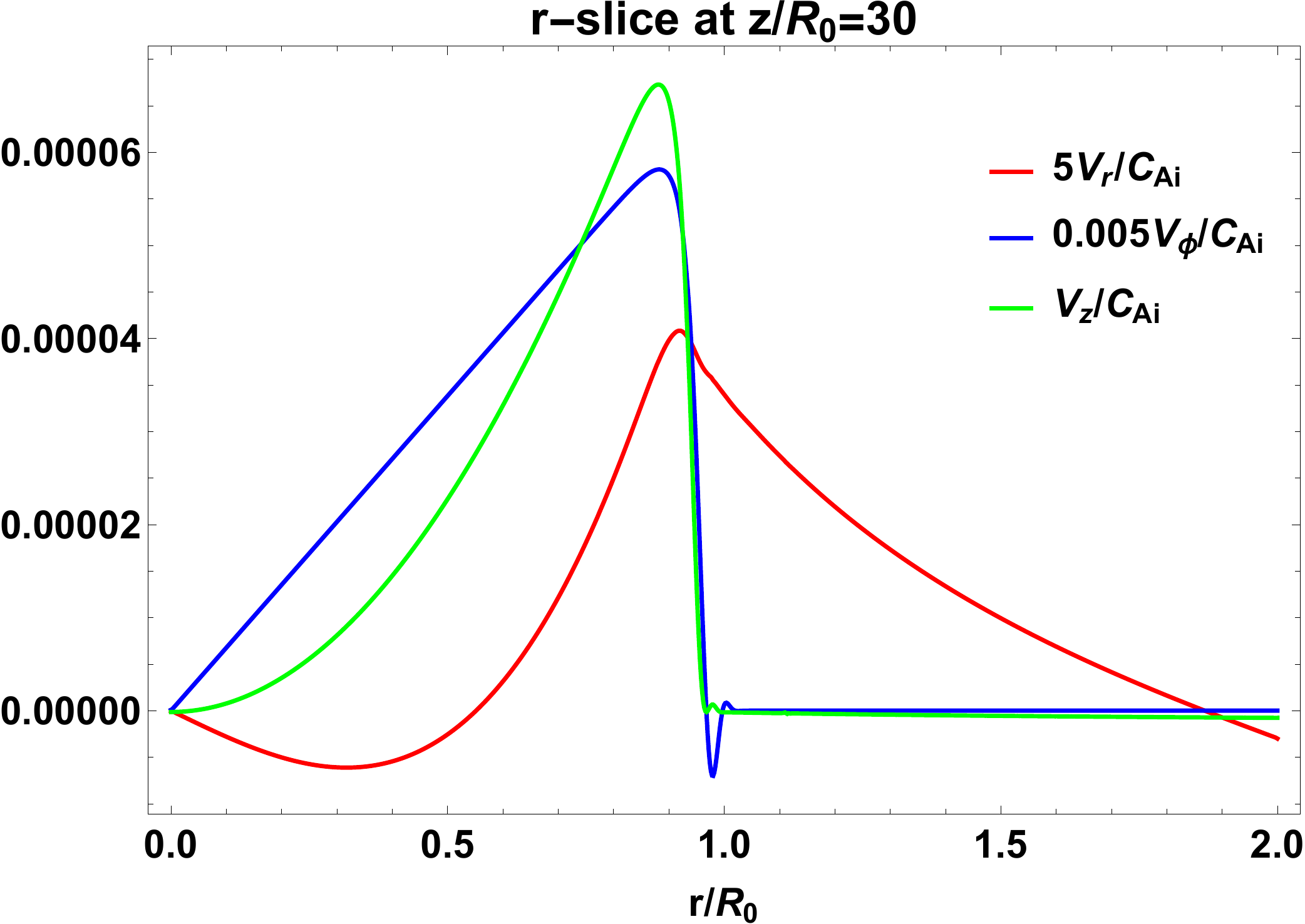}
	\caption{Plasma velocity perturbations sliced along the radial coordinate (for Driver 1) at dimensionless time $t=10$.}
	\label{fig:ld_r_slice}
\end{figure}

Due to the fact that the characteristics of the mother Alfv\'en wave explicitly depend on the radial coordinate, it would be natural to expect the induced perturbations to \lq inherit\rq\ this dependence. Indeed, it can be seen that all induced perturbations depend on the radial coordinate, except density which seems to be radially uniform. To examine this uniformity and highlight other features of the simulations carried out in the present study, we take a few slices along the tube axis (see Figs. \ref{fig:ld_z_slices} and \ref{fig:ld_separate_z_slices}). Figure \ref{fig:ld_z_slices} shows the perturbations of density, axial, radial and azimuthal velocities sliced along the $z$-axis of a tube for different values of the radial coordinate (all slices are inside the tube). First of all, among the features highlighted in Fig. \ref{fig:ld_z_slices}, it could be noticed that the perturbations of density together with perturbations of axial and radial velocities propagate at the Aflv\'en speed as the mother Alfv\'en wave (perturbation of azimuthal velocity), reaching $z \approx 31.6$ by the time $t=10$.. Moreover, these perturbations are induced with the double of the frequency of the Alfv\'en wave. This feature has also been observed regarding standing waves \citep{2011A&A...526A..80V}. Also, the fast sausage wave \citep[see e.g.][]{Edwin1983,2003A&A...412L...7N,2014ApJ...781...92V,2019ApJ...886..112K, 2021MNRAS.505.3505K} can be observed in the right part of the plots ($z>31.6$) for curves corresponding to density and radial velocity perturbations. Looking at the density plot, signatures of a mass flow could be readily observed, which is not a concern of the present study. At all slices the perturbations of density have the same value, however, to show this explicitly, we plotted all axial slices of each parameter on separate plots of Fig.~\ref{fig:ld_separate_z_slices}. As can be seen from Fig.~\ref{fig:ld_separate_z_slices}, all $z$-slices for the density perturbations coincide, except for the left part of the plot ($z<12.9$) where slow waves are present \citep[see e.g.][for a recent study of the nonlinear coupling of Alfvén and slow magnetoacoustic
waves]{Ballester2023}. Thus, the density perturbation induced by the torsional Alfv\'en wave is radially uniform. For other plasma parameters shown in Fig. \ref{fig:ld_separate_z_slices}, dependence on the radial coordinate is readily noticed. To visualize these dependencies, we plot an $r$-slice of the velocity perturbations in Fig.~\ref{fig:ld_r_slice}. As noticed from our setup, the perturbations of the azimuthal velocity increase linearly with the radius. At the same time, the induced perturbations of axial and radial velocities have parabolic and cubic dependencies on the radial coordinate, respectively, as shown in \citet{Shestov2017}.

Thus, the Alfv\'en wave driver in the entire flux tube volume (Driver 1) results in the induced density perturbation that is uniform in the radial direction. It is in contrast to the non-collective nature of the mother Alfv\'en wave and to the evident radial dependence of other induced plasma perturbations.
%In the succeeding subsection, we continue our simulation for various Alfv\'en drivers in order to reveal properties of the induced density perturbations.

\subsection{\lq\lq Annulus\rq\rq\ driver (Driver 2)}
\begin{figure*}
	\includegraphics[width=\textwidth]{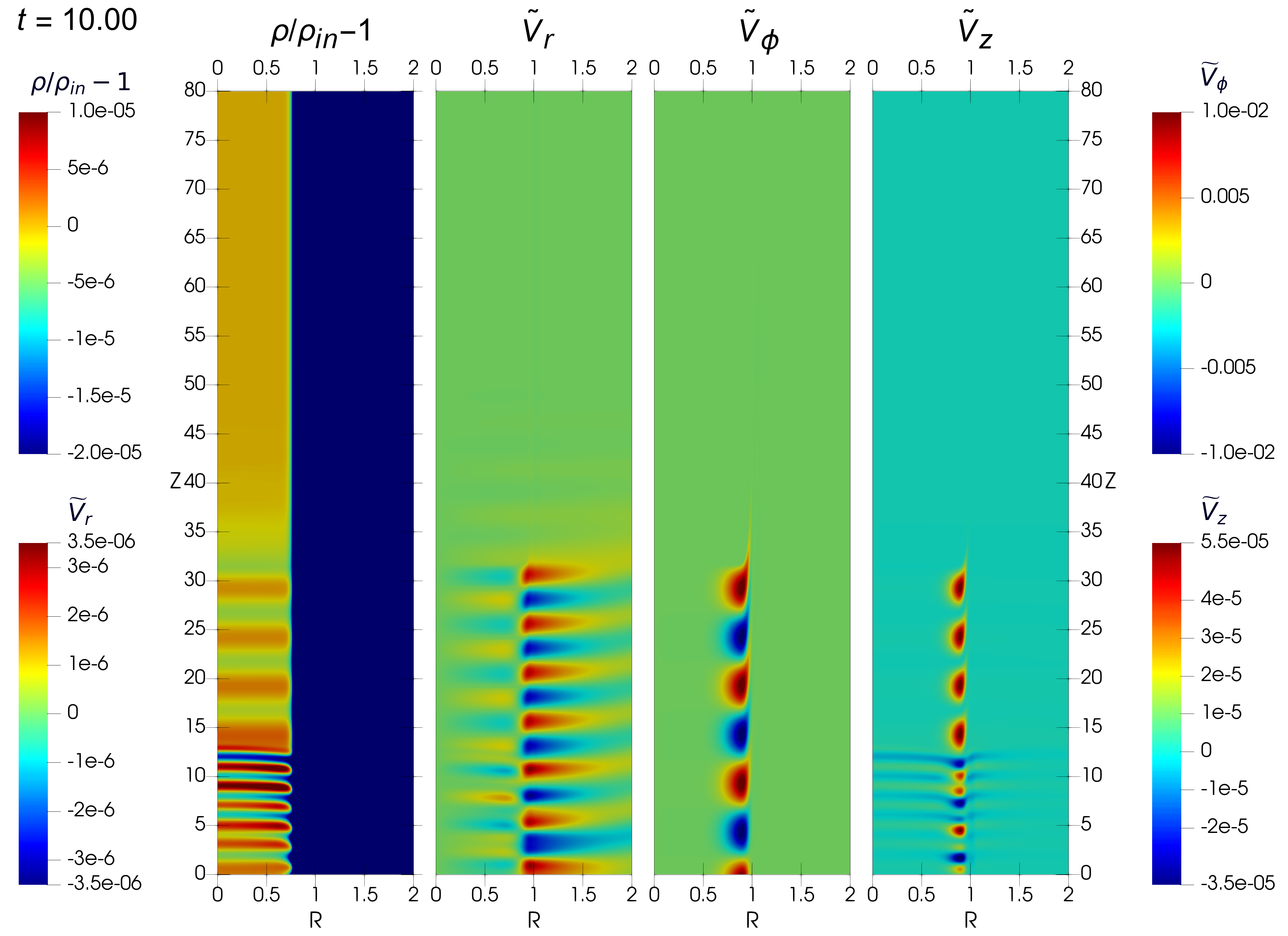}
	\caption{Dimensionless perturbations of density, axial and radial velocities induced by a torsional Alfv\'en wave (perturbation of azimuthal velocity) existing only in a part of the tube volume (i.e. with Driver 2, see Fig.~\ref{fig:setup}.b) at dimensionless time $t=10$. Due to axial symmetry, only one half of the magnetic flux tube is shown. The full animation can be seen in supplementary material.}
	\label{fig:border_full}
\end{figure*}
 Figure \ref{fig:border_full} shows the propagation of the torsional Alfv\'en wave, driven by Eqs.~(\ref{v_driver}) and (\ref{b_driver}) with subscript 2 (i.e. by Driver 2). In this case, the Alfv\'en wave exists only in a narrow annulus near the boundary of the magnetic flux tube (see the blue curve in Fig. \ref{fig:setup}.b). Figure \ref{fig:border_full} demonstrates that despite such a localised structure of the mother Alfv\'en wave driver, the induced perturbation of density is still radially uniform. In turn, the perturbations of axial and radial velocities become more narrow in accordance with the structure of the chosen Alfv\'en driver. It is worth noting that other features like phase-mixing and the existence of fast and slow waves are also observed in this simulation.

\begin{figure*}
	\includegraphics[width= \textwidth]{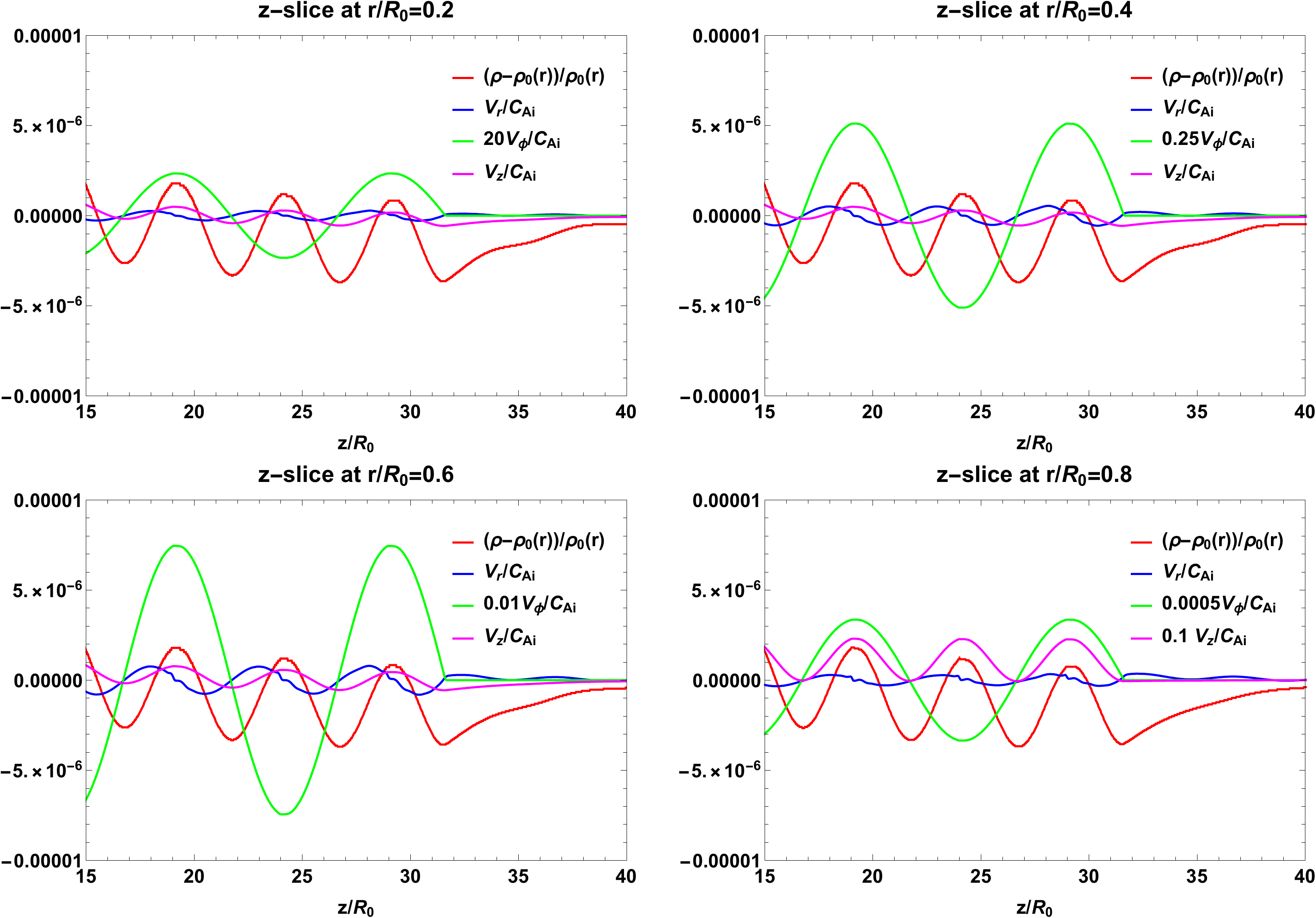}
	\caption{Plasma parameters sliced along the $z$-axis for different values of the radial coordinate, perturbed by the mother Alfv\'en wave driven by Driver 2 (see Fig.~\ref{fig:border_full}) at dimensionless time $t=10$. (some parameters have been re-scaled for visualizing purposes).}
	\label{fig:bd_z_slices}
\end{figure*}

\begin{figure*}
	\includegraphics[width= \textwidth]{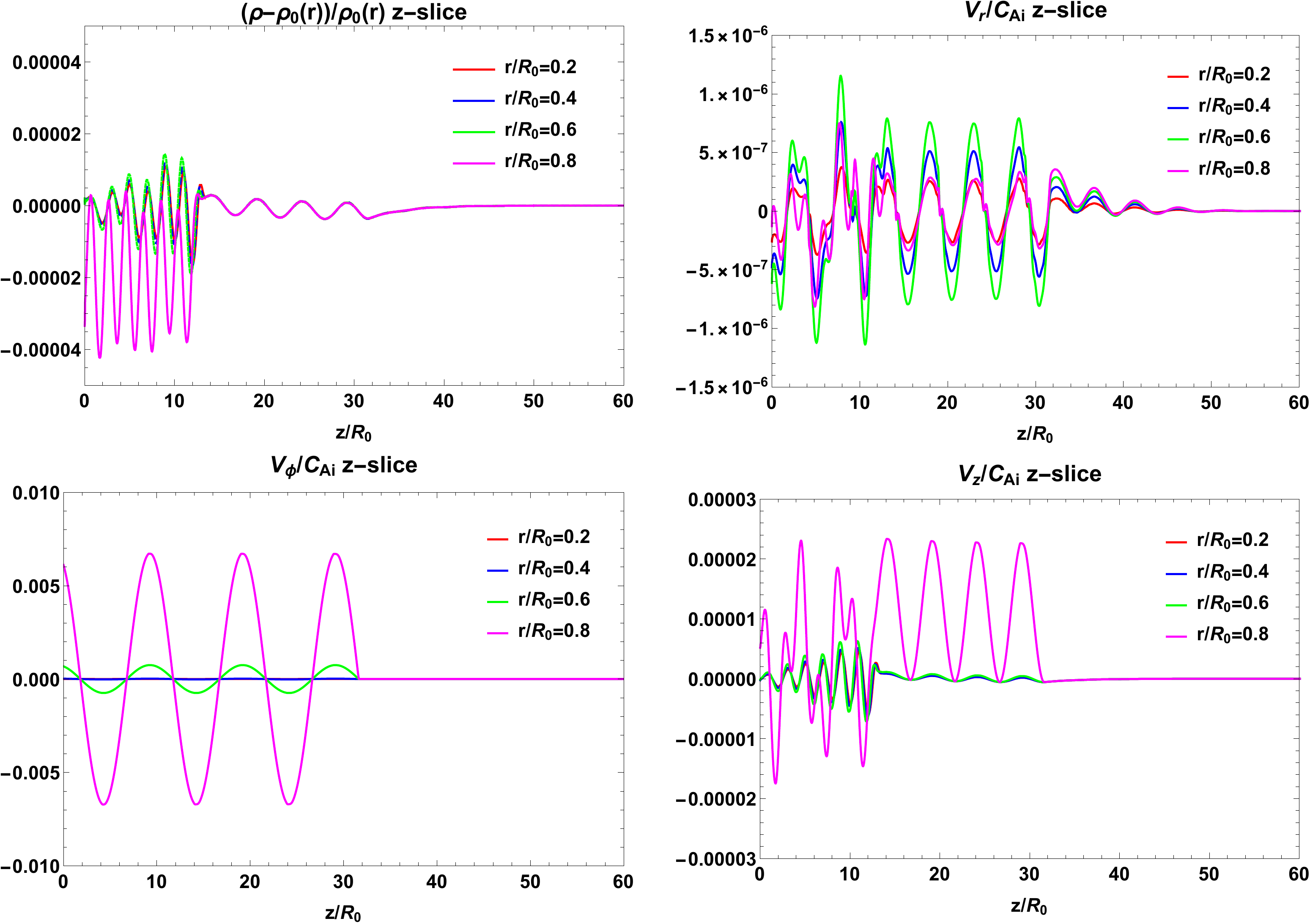}
	\caption{Comparison of the plasma parameters sliced along the $z$-axis for different values of the radial coordinate (for Driver 2) at dimensionless time $t=10$. Each plot compares axial slices of one parameter taken at different values of the radial coordinate.}
	\label{fig:bd_separate_z_slices}
\end{figure*}

\begin{figure}
	\includegraphics[width= \columnwidth]{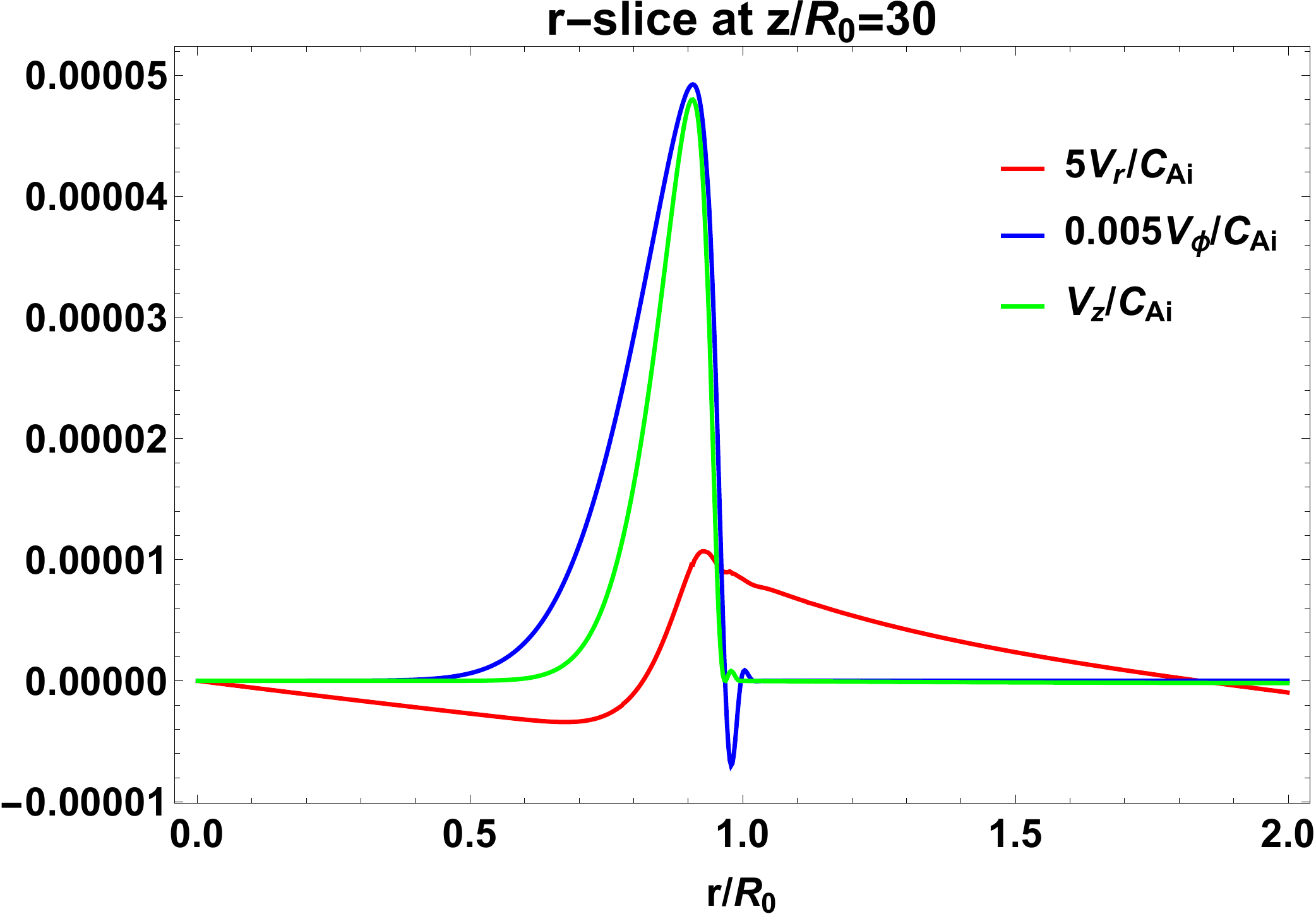}
	\caption{Plasma velocity perturbations sliced along the radial coordinate (for Driver 2) at dimensionless time $t=10$.}
	\label{fig:bd_r_slice}
\end{figure}

One-dimensional $z$-slices of the plasma parameters plotted in Fig.~\ref{fig:bd_z_slices} and Fig.~\ref{fig:bd_separate_z_slices} prove that the density perturbation remains uniform in the radial direction, while other perturbations actually depend on the radial coordinate because of the radial dependence of the mother Alfv\'en wave. Radial profiles of the velocity perturbations are shown in Fig.~\ref{fig:bd_r_slice}.

The results of both simulations (Figs.~\ref{fig:linear_full} and \ref{fig:border_full}) show that the induced plasma perturbations propagate with a double frequency of the mother Alfv\'en wave, as expected. On the other hand, all induced perturbations reveal dependencies on the radial coordinate, except the perturbation of density, which is radially uniform independently of the radial profile of the Alfv\'en driver. This numerical finding suggests that the mechanisms that induce density perturbations and perturbations of other plasma parameters (e.g. axial velocity) may be different, which we address in detail in Sec.~\ref{s:Analytics}.

\section{Analytical interpretation}\label{s:Analytics}
To analytically describe how torsional Alfv\'en waves induce compressive perturbations, we use the model of the straight magnetic cylinder and take into account perturbations up to the second order of smallness. Moreover, we assume that there are no compressive perturbations in the first order as they arise due to the mother Alfv\'en wave. Also, we concentrate on Alfv\'en waves propagating only in one direction. Under these assumptions, we can obtain equations describing the dynamics of the axial velocity and density perturbations inside and outside the magnetic flux tube from the set of Eqs.~(\ref{Cont_2nd_order})--(\ref{State_2nd_order}),
\begin{multline}
	\left[\left(C_{A_{i,e}}^2+C_{S_{i,e}}^2\right)\left(\frac{1}{r}\pder{}{r}\left(r\pder{}{r}\right)\right)\hat{D}_{T_{i,e}}-\hat{D}_{A_{i,e}}\hat{D}_{S_{i,e}}\right]\tilde{v}_z=\\-\frac{C_{A_{i,e}}^2}{8\pi\rho_{0_{i,e}}}\frac{1}{r}\pder{}{r}\left(r\pder{}{r}\right)\frac{\partial^2 \tilde{B}_\phi^2}{\partial t\partial z},
	\label{vz_eq}
\end{multline}
\begin{equation}
	\left[\left(C_{A_{i,e}}^2+C_{S_{i,e}}^2\right)\left(\frac{1}{r}\pder{}{r}\left(r\pder{}{r}\right)\right)\hat{D}_{T_{i,e}}-\hat{D}_{A_{i,e}}\hat{D}_{S_{i,e}}\right]\tilde{\rho}=0,
	\label{density_eq}
\end{equation}
where we have $\hat{D}_j = \partial^2/\partial t^2 - C_j^2\partial^2/\partial z^2$. We note that $C_A$, $C_S$, and $C_T$ are respectively the Alfv\'en, sound, and tube speeds. Subscripts i and e correspond to the internal and external media, respectively. Eq.~(\ref{vz_eq}) describes how torsional Alfv\'en waves induce perturbations of the axial plasma velocity by a ponderomotive force on the RHS due to the presence of the term proportional to the magnetic pressure perturbation in the Alfv\'en wave. {In contrast to the thin flux tube approximation \citep{2011A&A...526A..80V, 2012A&A...544A.127V}, in the considered full MHD model this ponderomotive term depends on the derivatives in both the axial and radial directions \citep[see also][]{Nakariakov1997}. Apparently, this radial dependence of the ponderomotive force is responsible for the radial structure of the perturbation.} On the other hand, there is no such term on the RHS of Eq.~(\ref{density_eq}) for density perturbations. This means that the radially uniform density perturbations seen in our numerical simulations in Figs.~\ref{fig:linear_full} and \ref{fig:border_full} are induced indirectly via driving the axial (and apparently radial) velocity by the mother Alfv\'en wave.
We also note that this model (Eqs.~(\ref{Cont_2nd_order})--(\ref{State_2nd_order}) and Eq.~(\ref{vz_eq})) coincides with that from \citet{Scalisi_2021}. A more general model, accounting for the effect of thermal misbalance, was considered by \citet{Belov2022}, in the long-wavelength limit and using the thin flux tube approximation.

To find out what exactly causes those collective density perturbations inside the magnetic flux tube, we concentrate on perturbations of the form $F\left(r, \xi=z-C_{A_i}t\right)$ and rewrite Eq.~(\ref{vz_eq}) as
\begin{equation}
	\frac{1}{r}\pder{}{r}\left(r\pder{}{r}\right)\pderN{}{\xi}{2}\left[\tilde{v}_z-\frac{\tilde{B}_\phi^2}{8\pi\rho_{0_i} C_{A_i}}\right]=0.
	\label{vz_eq_simplified}
\end{equation}
Eq.~(\ref{vz_eq_simplified}) coincides with Eq.~(28) from \citet{Scalisi_2021}, where its solution was found in the form $\tilde{v}_z=\tilde{B}_\phi^2/8\pi\rho_{0_i} C_{A_i}$. In the present study, we provide a more general solution by integrating Eq.~(\ref{vz_eq_simplified}) four times,
\begin{equation}
	\tilde{v}_z=\frac{\tilde{B}_\phi^2}{8\pi\rho_{0_i}  C_{A_i}}+A_1\left(\xi\right)+A_2\left(\xi\right)\ln\left(r\right)+A_3\left(r\right)+A_4\left(r\right)\xi,
	\label{vz_full_sol}
\end{equation}
where $A_j$ are arbitrary functions which should be determined from the boundary conditions. We set $A_2$ and $A_4$ to zero because our solution should be limited when $r=0$ and $\xi\rightarrow\pm\infty$, respectively. Also, since we assume that $\tilde{v}_z$ is induced by the Alfv\'en wave, then there should be no perturbations ahead of the Alfv\'en wave front. This means that we have $\left.\tilde{v}_z\right|_{\xi=0}=0$ for any $r$. Thus, we also set $A_3$ to zero. As a result, we can rewrite our solution (\ref{vz_full_sol}) as
\begin{equation}
	\tilde{v}_z=\frac{\tilde{B}_\phi^2}{8\pi\rho_{0_i} C_{A_i}}+V\left(\xi\right).
	\label{vz_trunc_sol}
\end{equation}
Here, the arbitrary function $V\left(\xi\right)$ should be determined from the condition of continuity of the total pressure and radial velocity at the tube boundary. Next, using Eqs.~(\ref{Cont_2nd_order})--(\ref{State_2nd_order}) together with solution (\ref{vz_trunc_sol})  we obtain expressions for other plasma perturbations induced by the torsional Alfv\'en wave inside the magnetic flux tube,
\begin{align}
	&\tilde{\rho}=\frac{\rho_{0_i} C_{A_i}}{C_{S_i}^2}V,\label{solutions_rho}\\
    &\tilde{P}_T=\frac{\rho_{0_i} C_{A_i}^3}{C_{S_i}^2}V,\label{solutions_pt}\\
	%&\tilde{B}_z=\frac{B_{0_i}}{C_{A_i}}\left[\left(\frac{C_{A_i}^2-C_{S_i}^2}{C_{S_i}^2}\right)V-\frac{\tilde{B}_\phi^2}{8\pi\rho_{0_i} C_{A_i}}\right],\\
	&\tilde{v}_r=\left(\frac{C_{A_i}^2-C_{S_i}^2}{C_{S_i}^2}\right)\frac{r}{2}\deriv{V}{\xi}-\frac{1}{8\pi\rho_{0_i} C_{A_i}}\frac{1}{r}\int_{0}^{r}{r\pder{\tilde{B}_\phi^2}{\xi}dr}.
	\label{solutions_vr}
\end{align}
From Eqs.~(\ref{solutions_rho})--(\ref{solutions_vr}), it could be noticed that the density perturbation is proportional to $V\left(\xi\right)$ and it does not directly depend on the mother Alfv\'en wave. Due to this fact, the perturbation of density is radially uniform which arises independently of the mother Alfv\'en wave and its radial profile. This coincides with the numerical results obtained in Sec.~\ref{s:Num_res}. The physical meaning of the function $V\left(\xi\right)$ can be understood from Eqs.~(\ref{vz_trunc_sol}) and (\ref{solutions_vr}). It is an additional axial velocity arising due to the interaction with the external medium through the perturbation of the tube boundary.

The mechanism of the density perturbation induction is secondary to the mechanism of the ponderomotive force \citep{1999JPlPh..62..219V} which roots in the radial and axial velocity perturbations by the mother Alfv\'en wave. This secondary mechanism {can be explained qualitatively with a concept of} a non-equal elasticity of a magnetic flux tube in the radial and axial directions, respectively. More specifically, the mother Alfv\'en wave induces radial and axial velocity perturbations which stretch the magnetic flux tube in the corresponding directions. Moreover, in the radial direction, the tube boundary experiences a reaction from the external medium, thus making the resulting density perturbations to depend on the boundary conditions. {The exact identification of the parameters describing non-equal elasticity has not been conducted here and is subject to the next study.}

Let us illustrate this dependence on the conditions at the tube boundary using two limiting cases. If the interaction with the external medium is absent, i.e. the perturbation of the total pressure $\tilde{P}_T$ is zero \citep[see e.g. the so-called thin flux tube approximation,][]{1996PhPl....3...10Z}, it follows from Eq.~(\ref{solutions_pt}) that $V=0$. Using this and Eq.~(\ref{solutions_rho}), the perturbation of density is not induced.
Now consider the case that the  magnetic flux tube possesses rigid walls, i.e. perfectly reflecting boundaries with
$\left.\tilde{v}_r\right|_{r=R_0}=0$. Then, we can readily determine $V\left(\xi\right)$ from Eq.~(\ref{solutions_vr}) as
\begin{equation}
	V=\frac{C_{S_i}^2}{4\pi\rho_{0_i} C_{A_i}\left(C_{A_i}^2-C_{S_i}^2\right)}\frac{1}{R_0^2}\int_{0}^{R_0}{r\tilde{B}_\phi^2dr}.
	\label{V_simple}
\end{equation}
It is clearly seen from Eq.~(\ref{V_simple}) that $V\left(\xi\right)$ and the corresponding perturbations of density $\tilde{\rho}$ (\ref{solutions_rho}) are determined by the interplay between the mother Alfv\'en wave and the rigid tube boundary which represents the external medium in this case, and do not depend on the radial coordinate, $r$.

In the general case, both conditions set for radial velocity and total pressure at the tube boundary should be taken into account.
In this case, $V\left(\xi\right)$ can be determined via modified Bessel functions and an effective wavenumber in the radial direction \citep[see e.g.][for the case of field free external medium and sinusoidal Aflv\'en driver]{Andreassen1983}.
{The perturbation theory developed here is applicable for all $\beta$, except the case $C_{A_i}=C_{S_i}$ in which the perturbation amplitudes become singular (see e.g. Eq.~(\ref{V_simple})).}

\section{Summary and conclusions}\label{s:Conclusions}
An advanced understanding of the processes connected to the propagation of torsional Alfv\'en waves in solar coronal plasma structures is crucial not only for the problems of coronal heating and solar wind acceleration but also for the revealing of characteristic observational signatures of Alfv\'en waves. In this regard, we have studied the collectivity of the compressive perturbations induced by the torsional Alfv\'en waves in a loop manifested through a uniform radial distribution. Both analytical and numerical approaches have been used in this study.

First of all, the numerical simulation of propagating torsional Alfv\'en waves in a cylindrical magnetic flux tube has been performed for different Alfv\'en wave drivers: when the Alfv\'en wave exists in the entire volume of a magnetic flux tube and when it is localized in a narrow annulus near the tube boundary. It was found that for both drivers, the induced density perturbations are radially uniform, unlike the perturbations of axial and radial velocities which depend on the radial coordinate as prescribed by the mother Alfv\'en wave.

To interpret our numerical findings,  we have developed the analytical model of a straight magnetic cylinder with perturbations up to the second order of smallness. For the mother Alfv\'en wave propagating in one direction, we have obtained the ponderomotive force term in Eq.~(\ref{vz_eq}) describing the axial velocity perturbations, while this term is absent in Eq.~(\ref{density_eq}) for density perturbations. Based on this and from the analysis performed, we conclude that the mechanism of the loop density perturbation is secondary to the ponderomotive force mechanism, which arises due to the interaction of the radial velocity perturbations with the external medium at the loop boundary. In other words, it can be seen as a non-equal elasticity of the magnetic flux tube in the radial and axial directions, due to which the appearance of radially uniform density perturbations in torsional Alfv\'en waves strongly depends on the tube boundary conditions. We have provided an explicit analytical illustration of this dependence in two limiting cases, which are the loop with no perturbations of the total pressure at the boundary (i.e. no interaction with the external medium, used in e.g. the thin flux tube approximation), and the loop with perfectly reflecting rigid boundaries (i.e. no perturbation of the radial velocity). In the former case, there should be no induced density perturbations, while in the latter, the perturbation of density is explicitly shown to have no dependence on the radial coordinate, see Eqs.~(\ref{solutions_rho}) and (\ref{V_simple}).
In the general case, the continuity of both the total pressure and radial velocity should be taken into account.

Hence, we have demonstrated numerically and analytically that nonlinear torsional Alfv\'en waves can produce radially uniform density perturbations independently of the radial profile of the mother Alfv\'en wave. Based on this finding, we can conclude that this perturbation is collective, in contrast to the perturbations of axial and radial velocities whose radial structure is prescribed by that of the mother Alfv\'en wave and which are therefore non-collective. The existence of a collective and radially uniform density perturbation accompanying torsional Alfv\'en waves could provide an additional observational signature of Alfv\'en waves in the upper layers of the solar (and stellar) atmosphere.

Since this effect is nonlinear, the amplitude of the induced density perturbations is proportional to the amplitude of the driver squared $A_0^2$ and the plasma parameter $\beta$ as $A_0^2/[4\pi(1-\beta)]$ (see Eqs.~(\ref{solutions_rho}) and (\ref{V_simple})). {We restrict our study to the regime with $\beta<1$, typical for the solar corona. In the opposite case of $\beta>1$, we expect to see the slow wave propagating ahead the Alfv\'en wave. Hence, the compressive perturbations induced by the Alfv\'en waves are not likely to be observed alone due to the presence of the slow wave \citep[see e.g.][for a 1D illustration]{Boynton1996}}.  Thus, {for the amplitude of the Alfv\'en driver 1\% of the local Alfv\'en speed} and $\beta=0.2$ used in this work to avoid other higher-order nonlinear effects, the amplitude of the density perturbation was found to be just a fraction of a percent. However, for the Alfv\'en wave {with the amplitude 20\% of the local Alfv\'en speed} and $\beta=0.6$ \citep[see e.g.][]{Tsiklauri2002}, one should expect to obtain the density perturbation amplitude about 1\%. For example, MHD-wave-caused low-amplitude perturbations of the coronal plasma density have been previously shown to effectively modulate the observed dynamic spectra of solar radio bursts of various types \citep[see e.g.][]{2006SoPh..237..153K, 2013A&A...550A...1K, Kolotkov2018}. More specifically, \citet{Kolotkov2018} showed that the density perturbation with a relative amplitude of about 1\% and propagating at about the local Alfv\'en speed in the corona (about 0.7 solar radii above the surface) can cause up to 200\% modulation of the low-frequency plasma emission intensity. Likewise, the microwave gyrosynchrotron emission is known to be highly sensitive to low-amplitude variations of the coronal plasma parameters \citep[see e.g.][for recent studies]{2022MNRAS.516.2292K, 2022ApJ...937L..25S}. Nonetheless, the forward modelling accounting for a specific emission mechanism and the response function of a specific observational instrument is needed to address this question properly, which may constitute an interesting follow-up of this study.

\section*{Acknowledgements}

The study was supported in part by the Ministry of Science and Higher Education of Russian Federation under State assignment to educational and research institutions under Project No. FSSS-2023-0009 and No. 0023-2019-0003.
The work of DYK and VVB was supported by the Latvian Council of Science Project \lq\lq Multi-Wavelength Study of Quasi-Periodic Pulsations in Solar and Stellar Flares\rq\rq\ No. lzp-2022/1-0017.
%%%%%%%%%%%%%%%%%%%%%%%%%%%%%%%%%%%%%%%%%%%%%%%%%%
\section*{Data Availability}

The data underlying this article will be shared on reasonable request
to the corresponding author.



\bibliographystyle{mnras}
\bibliography{refs} % if your bibtex file is called example.bib

\appendix
\section{Governing equations}\label{App_A}
\begin{equation}
	\pder{\tilde{\rho}}{t}+\frac{1}{r}\rho_0\pder{r\tilde{v}_r}{r}+\rho_0\pder{\tilde{v}_z}{z}=0,
	\label{Cont_2nd_order}
\end{equation}
%%%%%%%%%%%%%%%%%%%%%%%%%%%%%%%%%
\begin{multline}
	\rho_0\pder{\tilde{v}_r}{t}-\frac{1}{r}\rho_0\tilde{v}_\phi^2=-\pder{\tilde{P}}{r}-\\-\frac{1}{4\pi}\left(\frac{1}{r}\tilde{B}_\phi\pder{r\tilde{B}_\phi}{r}-B_0\left(\pder{\tilde{B}_r}{z}-\pder{\tilde{B}_z}{r}\right)\right),
	\label{Motion_r_2nd_order}
\end{multline}
%%%%%%%%%%%%%%%%%%%%%%%%%%%%%%%%%
\begin{equation}
	\rho_0\pder{\tilde{v}_\phi}{t}=\frac{B_0}{4\pi}\pder{\tilde{B}_\phi}{z},
	\label{Motion_phi_2nd_order}
\end{equation}
%%%%%%%%%%%%%%%%%%%%%%%%%%%%%%%%%
\begin{equation}
	\rho_0\pder{\tilde{v}_z}{t}=-\pder{\tilde{P}}{z}-\frac{\tilde{B}_\phi}{4\pi}\pder{\tilde{B}_\phi}{z},
	\label{Motion_z_2nd_order}
\end{equation}
%%%%%%%%%%%%%%%%%%%%%%%%%%%%%%%%%
\begin{equation}
	\pder{\tilde{B}_r}{t}-B_0\pder{\tilde{v}_r}{z}=0,
	\label{Induction_r_2nd_order}
\end{equation}
%%%%%%%%%%%%%%%%%%%%%%%%%%%%%%%%%
\begin{equation}
	\pder{\tilde{B}_\phi}{t}-B_0\pder{\tilde{v}_\phi}{z}=0,
	\label{Induction_phi_2nd_order}
\end{equation}
%%%%%%%%%%%%%%%%%%%%%%%%%%%%%%%%%
\begin{equation}
	\pder{\tilde{B}_z}{t}+\frac{1}{r}B_0\pder{r\tilde{v}_r}{r}=0,
	\label{Induction_z_2nd_order}
\end{equation}
%%%%%%%%%%%%%%%%%%%%%%%%%%%%%%%%%
\begin{equation}
	\frac{1}{r}\pder{r\tilde{B}_r}{r}+\pder{\tilde{B}_z}{z}=0,
	\label{Div_2nd_order}
\end{equation}
%%%%%%%%%%%%%%%%%%%%%%%%%%%%%%%%%
\begin{equation}
	C_{V}\rho_0\pder{\tilde{T}}{t}-
	\frac{k_{\mathrm{B}}T_0}{m}\pder{\tilde{\rho}}{t}=0,
	\label{Energy_2nd_order}
\end{equation}
%%%%%%%%%%%%%%%%%%%%%%%%%%%%%%%%%%
\begin{equation}
	\tilde{P}=\frac{\mathrm{k_B}}{m}\left(\rho_0 \tilde{T}+\tilde{\rho} T_0\right),
	\label{State_2nd_order}
\end{equation}

%\begin{thebibliography}{99}
%\end{thebibliography}

% Don't change these lines
\bsp	% typesetting comment
\label{lastpage}
\end{document}